\def\Rb{$^{87}$Rb }

\def\ket#1{\left|#1\right\rangle}
\def\ketF#1{\left(#1\right)}
\def\ev#1{\left\langle#1\right\rangle}
\newcommand{\Jeffop}{\hat{J}_{\rm eff}}

\documentclass[aps,prl,twocolumn,floatfix,10pt]{revtex4-1} 
\usepackage{graphicx,amsmath,amssymb,ulem}
\usepackage{epstopdf}
\usepackage{color}

\begin{document}

\title{Detecting multiparticle entanglement of Dicke states}

\author{Bernd L\"u{}cke$^1$, Jan Peise$^1$, Giuseppe Vitagliano$^2$, Jan Arlt$^3$, Luis Santos$^4$, G\'e{}za T\'o{}th$^{2,5,6}$, Carsten Klempt$^1$ }

\affiliation{$^1$Institut f\"ur Quantenoptik, Leibniz Universit\"at Hannover, Welfengarten~1, D-30167~Hannover, Germany}
\affiliation{$^2$Department of Theoretical Physics, University of the Basque Country UPV/EHU, P.O. Box 644, E-48080 Bilbao, Spain}
\affiliation{$^3$QUANTOP, Institut for Fysik og Astronomi, Aarhus Universitet, 8000 \AA{}rhus C, Denmark}
\affiliation{$^4$Institut f\"ur Theoretische Physik, Leibniz Universit\"at Hannover, Appelstra\ss{}e~2, D-30167~Hannover, Germany}
\affiliation{$^5$IKERBASQUE, Basque Foundation for Science, E-48011 Bilbao, Spain}
\affiliation{$^6$Wigner Research Centre for Physics, H. A. S., P.O. Box 49, H-1525 Budapest, Hungary}

\date{\today}

\begin{abstract}
Recent experiments demonstrate the production of many thousands of neutral atoms entangled in their spin degrees of freedom. We present a criterion for estimating the amount of entanglement based on a measurement of the global spin. It outperforms previous criteria and applies to a wide class of entangled states, including Dicke states. Experimentally, we produce a Dicke-like state using spin dynamics in a Bose-Einstein condensate. Our criterion proves that it contains at least genuine 28-particle entanglement. We infer a generalized squeezing parameter of $-11.4(5)$~dB.

\end{abstract}

\maketitle
Entanglement, one of the most intriguing features of quantum mechanics, is nowadays a key ingredient for many applications in quantum information science \cite{RevModPhys.81.865,guhne2009entanglement}, quantum simulation \cite{kim2010quantum,simon2011quantum} and quantum-enhanced metrology \cite{giovannetti2011advances}. 
Entangled states with a large number of particles cannot be characterized via full state tomography~\cite{Paris2004}, which is routinely used in the case of photons~\cite{White1999,2014arXiv1401.7526S}, 
trapped ions~\cite{Haffner2005},
 or superconducting circuits~\cite{neeley2010generation,dicarlo2010preparation}. A reconstruction of the full density matrix is hindered and finally prevented by the exponential increase of the required number of measurements. Furthermore, it is technically impossible to address all individual particles or even fundamentally forbidden if the particles occupy the same quantum state. Therefore, the entanglement of many-particle states is best characterized by measuring the expectation values and variances of the components of the collective spin ${\mathbf J} = (J_x, J_y, J_z)^T =\sum_i {\mathbf s}_i$, the sum of all individual spins ${\mathbf s}_i$ in the ensemble. 

In particular, the spin-squeezing parameter $\xi^2=N \frac{(\Delta J_z)^2}{\langle J_x\rangle^2+\langle J_y\rangle^2}$ defines the class of spin-squeezed states for $\xi^2<1$. This inequality can be used to verify the presence of entanglement, 
since all spin-squeezed states are entangled~\cite{Sorensen2001}. Large clouds of entangled neutral atoms are typically prepared in such spin-squeezed states, as shown in thermal gas 
cells~\cite{Hald1999}, at ultracold temperatures~\cite{Schleier-Smith2010,Chen2011,Sewell2012} and in Bose-Einstein condensates~\cite{Gross2010,Riedel2010,Hamley2012}.

\begin{figure}[t!]
		\includegraphics[width={\columnwidth}]{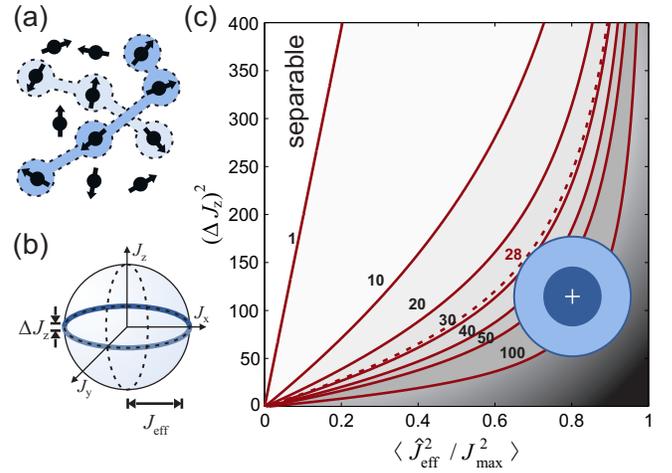}
		\caption{Measurement of the entanglement depth for a total number of $8000$ atoms. (a) The entanglement depth is given by the number of atoms in the largest non-separable subset (shaded areas). (b) The spins of the individual atoms add up to the total spin $\mathbf{J}$ whose possible orientations can be depicted on the Bloch sphere. Dicke states are represented by a ring around the equator with an ultralow width $\Delta J_z$ and a large radius $J_{\mathrm{eff}}$.  (c) The entanglement depth in the vicinity of a Dicke state can be inferred from a measurement of these values. The red lines indicate the boundaries for various entanglement depths. The experimental result is shown as blue uncertainty ellipses with one and two standard deviations, proving an entanglement depth larger than $28$ (dashed line). }
	\label{fig1}
\end{figure}

Systems with multiple particles may exhibit more than pairwise entanglement. Multiparticle
entanglement is best quantified by means of the so-called entanglement depth, defined as the number of particles in the largest non-separable subset [see Fig.~1 (a)]. There have been numerous experiments 
detecting multiparticle entanglement involving up to $14$ 
qubits in systems, where the particles can be addressed individually~\cite{Sackett2000,Haffner2005,PhysRevLett.106.130506, Wieczorek2009,Prevedel2009,gao2010experimental}.
Large ensembles of neutral atoms pose the additional challenge of obtaining the entanglement depth from collective measurements. Following the criterion for $k$-particle entanglement of Ref.~\cite{Sorensen2001a}, multiparticle entanglement  
has been experimentally demonstrated in spin-squeezed Bose-Einstein condensates~\cite{Gross2010}. However, the method only applies to spin-squeezed states, which constitute a small subset of all possible entangled many-particle states. Moreover, the strong entanglement of states with extreme sub-shot-noise fluctuations is not detected under influence of minimal experimental noise~\cite{Suppl}. Whereas entanglement detection for more general entangled states has already been developed~\cite{Toth2007,Tura2013}, it is desirable to extend these methods towards the detection of multiparticle entanglement.

In this Letter, we introduce a method for the quantification of entanglement. Our criterion is applicable to both spin-squeezed and more extreme states, yielding superior results compared to the inspiring work by S\o{}rensen/M\o{}lmer~\cite{Sorensen2001a} and Duan~\cite{Duan2011}. It enables us to quantify the multiparticle entanglement of an experimentally created Dicke-like state, yielding a minimum entanglement depth of $28$. In addition, we extract a generalized squeezing parameter, which is also applicable to Dicke states, of $-11.4(5)$~dB, so far the best reported value in any atomic system.

Dicke states~\cite{Dicke1954} constitute a particularly relevant class of highly entangled, but not spin-squeezed states. They are simultaneous eigenstates $\ket{J,M}$ of ${\mathbf J}^2$ and $J_z$, and the spin-squeezing parameter $\xi^2$ does not detect them as entangled~\cite{Ma2011}. Nonetheless, Dicke states have optimal metrological properties~\cite{Krischek2011,Toth2012,Hyllus2012a} and can be used to reach Heisenberg-limited sensitivity~\cite{Holland1993}. They are also useful for quantum information processing tasks, such as $1\rightarrow(N-1)$ telecloning or open-destination teleportation~\cite{Chiuri2012}. Experimentally, high-fidelity Dicke states with small particle numbers have been created with photons~\cite{Wieczorek2009,Prevedel2009} and trapped ions~\cite{Haffner2005}, and have been detected by global measurements~\cite{Noguchi2012}.

Among other methods~\cite{Vanderbruggen2011,Bucker2011}, large numbers of atoms in Dicke states with $\ket{J,M=0}$ may be created in spinor Bose-Einstein condensates~\cite{Stamper-Kurn2013}. Spin dynamics creates a superposition of Dicke states with varying total number of particles in a process that resembles optical parametric down-conversion~\cite{Klempt2009,Klempt2010}. In previous work, the entanglement of these states was proven by a homodyne measurement~\cite{Gross2011} and by a test of the metrological sensitivity beyond shot noise~\cite{Luecke2011}. However, the achieved metrological sensitivity did not imply more than pair-wise entanglement~\cite{Hyllus2012a}.

\begin{figure}[t!]
			\includegraphics[width={\columnwidth}]{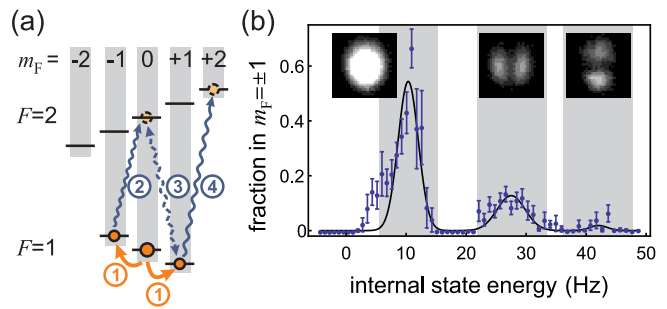}
	\caption{Preparation and detection of a Dicke-like state. (a) A Bose-Einstein condensate in the level $\ketF{F,m_F}=\ketF{1,0}$ generates clouds with the same number of atoms in the levels $\ketF{1,\pm 1}$ (1). A microwave pulse (2) transfers the atoms from $\ketF{1,-1}$ to $\ketF{2,0}$. Optionally, a microwave pulse (3) can be used to couple the two clouds for the measurement of $J_{\mathrm{eff}}$. Finally, the atoms in the level $\ketF{1,1}$ are transferred to $\ketF{2,2}$ before detection. (b) The number of atoms is measured by standard absorption imaging (insets). On well-resolved resonances depending on the internal state energy, distinct spatial modes are populated with a large fraction of the total number of atoms. The black line is a Gaussian fit to guide the eye. In our experiments, we use the resonance at $\approx 28$~Hz. }
	\label{fig2}
\end{figure}

For the generation of the desired Dicke states, we prepare a \Rb Bose-Einstein condensate of $2 \times 10^4$ atoms in a crossed-beam dipole trap with trapping frequencies of $2 \pi \times (200,150,150)$~Hz. Initially prepared in the Zeeman level $\ketF{F,m_F}=\ketF{1,0}$, atoms collide and form correlated pairs in the two Zeeman levels $\ketF{1,\pm1}$. These atoms are transferred to distinct spatial modes~\cite{Klempt2009,Scherer2010}, which are addressed by microwave dressing~\cite{Stamper-Kurn2013} the Zeeman level $\ketF{1,1}$ [Fig.~2 (b)]. In an experimental run, up to $N = 8 \times 10^3$ atoms are transferred to the first excited mode along the strongest trap axis within $240$~ms. Since they are transferred pairwise, we expect an equal number of atoms $N_{\pm 1}=\frac{N}{2}$ in the two Zeeman levels $\ketF{1,\pm 1}$. These atoms are highly entangled in analogy to optical parametric down-conversion. It is the central objective of this Letter to quantify the entanglement depth of the created many-particle state.

\begin{figure*}[t!]
			\includegraphics[width={2\columnwidth}]{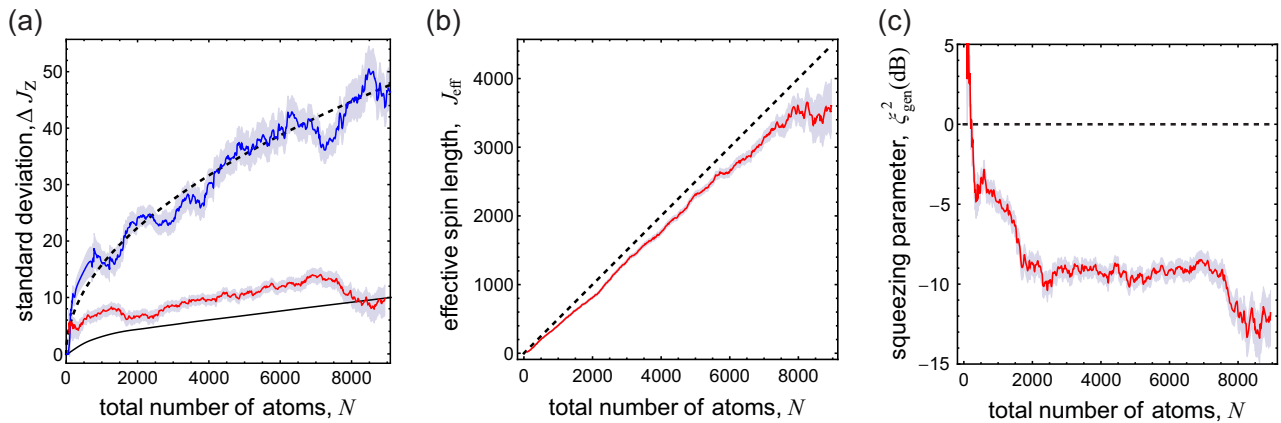}
	\caption{Characterization of the experimentally created Dicke-like state. (a)  Measurement of the width $\Delta J_z$ for varying total number of atoms (red line). Each value and its statistical uncertainty (gray shading) is calculated for a $1000$-atom interval within the total number of atoms. The measured values of $\Delta J_z$ are well below the shot noise limit (theory: black dashed line, experiment: blue solid line) and partially explained by a lower limit of the number dependent detection noise (black solid line). (b) The measured value of $J_{\mathrm{eff}}$ as a function of the total number of atoms almost reaches its optimal value (black dashed line). The inset shows that the normalized $J_{\mathrm{eff}}$ is slowly reduced during an additional hold time. (c) The recorded data allows for a determination of the optimal spin-squeezing parameter as a function of the total number of atoms. At a total of $8000$ atoms, it reaches a value of $-11.4(5)$~dB.}
	\label{fig3}
\end{figure*}

We restrict the description of the output state to the two relevant Zeeman levels $\ketF{1,\pm 1}$. In this 
pseudo-spin-$\frac{1}{2}$ system, we characterize the state by the collective spin $\mathbf{J}$, resulting from the sum of the individual pseudospins. In this picture, the ideal output state with equal number of atoms constitutes the Dicke state $\ket{J=\frac{N}{2}, M = 0}$ with vanishing fluctuations $\Delta J_z$.
The fluctuations of the collective spin can be measured directly by counting the number of atoms in the two Zeeman levels. For this purpose, we transfer the atoms to the levels $\ketF{2,0}$ and $\ketF{2,2}$ with microwave pulses [see Fig.~\ref{fig2} (a)]. Subsequently, the trap is switched off and a strong magnetic field gradient separates the spin components during ballistic expansion. The number of atoms is then measured by standard absorption imaging. The absolute number of atoms was calibrated~\cite{Luecke2011} and it was confirmed that shot noise fluctuations are observed for a coherent state 
 [see Fig.~3 (a)], which was created by splitting a Bose-Einstein condensate with a $\frac{\pi}{2}$ microwave pulse.

We measure $J_x$ and $J_y$ by rotating the total spin using a $\frac{\pi}{2}$ microwave coupling pulse on the $\ketF{1,1}$ to $\ketF{2,0}$ transition before the number measurement~[see Fig.~2 (a)]. Whether $J_x$ or $J_y$ is measured depends on the relation between the microwave phase and the phase of the initial Bose-Einstein condensate. The condensate phase represents the only possible phase reference in analogy to the local oscillator in optics. Intrinsically, it has no relation to the microwave phase, such that we homogeneously average over all possible phase relations in our measurements. For a given phase difference $\alpha$, a rotation yields a measurement of $J_\alpha \equiv \cos\alpha J_x + \sin\alpha J_y$. Averaging over all possible $\alpha$, the measured expectation value of
the second moment corresponds to $\frac{1}{2\pi}\int_0^{2\pi}{\langle J_\alpha^2 \rangle \mathrm{d}\alpha}={\left\langle  \frac{1}{2} (J_x^2 +J_y^2) \right\rangle}$. After a random rotation, we thus record the effective spin length $J_{\rm eff}^2=\langle \hat{J}_{\rm eff}^2\rangle =\langle J_x^2 +J_y^2\rangle $, which equals the spin length in the limit of vanishing $\ev{J_z^2}$~\footnote{For small particle numbers, ${J}_{\rm eff}$ is defined as {$J_{\rm eff}({J}_{\rm eff}+1)=\left\langle {J_x^2+J_y^2} \right\rangle$} with $0\le J_{\rm eff}\le\frac{N}{2}.$ For large particle numbers, we approximate {$J_{\rm eff}^2\approx\ev{J_x^2+J_y^2}$}.}. Dicke states can be ideally characterized by the measurement of a large $J_{\mathrm{eff}}$ and a small variance $(\Delta J_z)^2$
 [see Fig.~1 (b)]. 

Figure~3 (a) depicts the results of our measurement of $\Delta J_z$ depending on the total number of atoms $N$. The recorded fluctuations were corrected for the independently measured detection noise of $10.9(3)$ atoms to obtain the pure atomic noise. The detection noise was directly extracted from images of the detection beams and is mainly caused by the photoelectron shot noise on the camera. The measured atom number fluctuations are well below the atomic shot noise level, reaching down to $-12.4\pm1.2$~dB at a total number of $8000$~atoms. The fluctuations are almost independent of the total number of atoms with a small trend of $0.15 \sqrt{N}$. We do not record an increase of the measured fluctuations for a variable additional hold time of up to $420$~ms. Thus, we can exclude three-body losses, collisions with the background gas or radio-frequency noise as relevant noise sources. We attribute the measured fluctuations to an additional detection noise since photoelectron shot noise and the influence of technical noise of the imaging beams are expected to increase slightly for a larger number of atoms. The solid line in Fig. 3 (a) shows an estimated lower limit of this effect~\cite{Suppl}. 

A measurement of the effective spin length $J_{\rm eff}$ is presented in Fig.~3 (b). The values for $J_{\rm eff}$ almost reach their optimal value of $J_{\rm max}=\frac{N}{2}$. 
This measurement shows that the created state is nearly fully symmetric. After a variable hold time, the measured effective spin length diminishes slowly [see Fig.~3 (b), inset]. We thus conclude that the measurement result is limited by magnetic field gradients and collisions. Elastic collisions can transfer individual atoms to other spatial modes, reducing the ensemble's purity and the achievable effective spin length. The combined measurements of $\Delta J_z$ and $J_{\rm eff}$ prove that the created many-particle state is in the close vicinity of an ideal symmetric Dicke state.

The measurements can be combined to extract a generalized squeezing parameter $\xi^2_{\textrm{gen}}=(N-1) \frac{(\Delta J_z)^2}{\ev{J_x^2}+\ev{J_y^2}-N/2}$ which extends the concept of the spin-squeezing parameter to more general entangled states, including Dicke states~\cite{Vitagliano2011,Vitagliano2014,[{The new parameter has been used to study the dynamics of the modified Lipkin-Meshkov-Glick model in }][{}]Marchiolli2013}. Figure~3 (c) presents the measured generalized squeezing parameter as a function of the total number of atoms. Note that the quasi-constant plateau is not statistically significant. At a total of $N=8000$ atoms, it reaches a value of $-11.4(5)$~dB. This represents the best reported value reached in any atomic system.

\begin{figure}[t]
\includegraphics[width={1\columnwidth}]{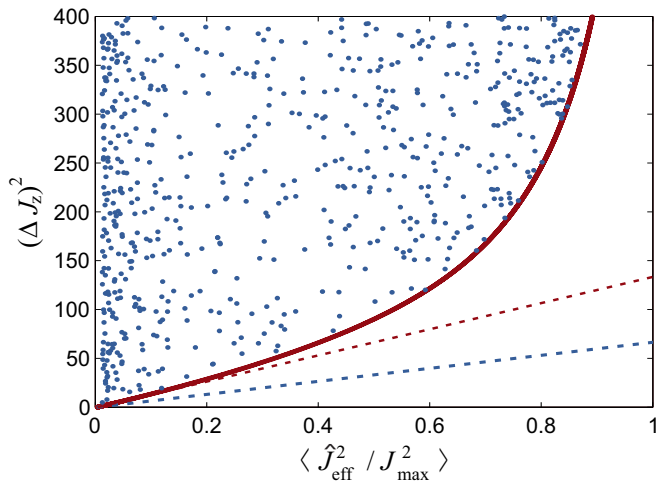}
	\caption{Detection of $k$-particle entanglement based on the total spin. The red line marks the boundary for $k$-particle entangled states with $N=8000$ and $k=28$ in the $(\ev{\Jeffop/J_{\rm max}^2}, (\Delta J_z)^2)$-plane. As a cross-check, random states with $k$-particle entanglement are plotted as blue dots, filling up the allowed region. The criterion of Ref.~\cite{Duan2011} only detects states that correspond to points below the dashed blue line. An improved linear criterion is gained from calculating a tangent to the new boundary (dashed red line).}
	\label{fig4}
\end{figure}

In addition to this proof of entanglement, the measured data allow for a quantification of the entanglement depth.
Given states with an entanglement depth $k$, it is possible to deduce a minimal achievable $(\Delta J_z)^2$ for each value of $J_{\mathrm{eff}}^2$~\cite{Suppl}. All states below this minimum must have an entanglement depth larger than $k$. 
It can be shown that the states on this boundary $\vert\Psi\rangle=\vert\psi\rangle^{\otimes \frac{N}{k}}$ are tensor products of identical $k$-particle states $\vert\psi\rangle$. Interestingly, these $k$-particle states are ordinary spin-squeezed states.  
Figure~4 shows the boundary in the case of $28$-particle entanglement at a total number of $8000$ atoms. As a cross-check, random states with $28$-particle entanglement are plotted in the figure. This confirms that our criterion is optimal and superior to the linear condition of Ref.~\cite{Duan2011}. Finally, the criterion detects a larger entanglement depth than the criterion given in Ref.~\cite{Sorensen2001a} when it is applied to spin-squeezed states with minimal experimental noise~\cite{Suppl}. It thus outperforms the original criterion in experimentally realistic situations. Beyond spin-squeezing, the criterion is applicable to unpolarized states and thus allows for an optimal evaluation of the entanglement depth of a Dicke-like state as created in our experiments.

Figure~1 (c) shows the entanglement depth of the created state for $8000$ atoms. The red lines present the newly derived boundaries for $k$-particle entanglement. All separable (unentangled) states are restricted to the far left of the diagram, as indicated by the $k=1$ line. The measured values of $(\Delta J_z)^2$ and $\langle \Jeffop^2 / J_{\rm max}^2\rangle$ are represented by uncertainty ellipses with one and two standard deviations. The center of the ellipses corresponds to an entanglement depth of $68$. With two standard deviations confidence, the data prove that our state has an entanglement depth larger than $28$. These numbers are only partly limited by the prepared state itself, but also by the number-dependent detection noise. This detection noise results in a larger measured value of $J_z^2$ and thus decreases the lower bound for the entanglement depth. This is the largest reported entanglement depth of Dicke-like states. In the future, the measured entanglement depth can be increased by an improved number detection, compensated magnetic field gradients and a faster spin dynamics.

In summary, we have presented a criterion for the detection of multi-particle entanglement based on a measurement of the ensemble's total spin. In the case of spin-squeezed states, the criterion outperforms the results of previous criteria in experimentally realistic situations. It also extends to more general entangled states, most importantly to Dicke states. We have applied the criterion to detect an entanglement depth larger than $28$ in an experimentally created Dicke-like state. The experimental results also allow for a determination of a generalized squeezing parameter of $-11.4(5)$~dB.

 We thank W. Ertmer, A. Smerzi, L. Pezz\'e and P. Hyllus for inspiring discussions. We acknowledge support from the Centre QUEST and from the DFG (Research Training Group 1729). We also thank the DFF, and the Lundbeck Foundation for support. We acknowledge support from the EMRP, which is jointly funded by the EMRP participating countries within EURAMET and the European Union. We thank the EU (ERC StG GEDENTQOPT, CHIST-ERA QUASAR), the MINECO (Project No. FIS2012-36673-C03-03), the Basque Government (Project No. IT4720-10), and the OTKA (Contract No. K83858).
\phantom{\cite{Stuart1994,Guehne2005}\cite{[{Note that the definition of $k$-particle entanglement
in this paper is equivalent to $k$-producibility as used in the literature.
See, for example, }][{. We use the less technical term {}``$k$-particle entanglement''
for the sake of simplicity.}]Guehne2005}\cite{Hyllus2012,Vitagliano}}

\bibliography{Luecke}

\hfill
\pagebreak

\setcounter{figure}{0}
\setcounter{table}{0}
\setcounter{equation}{0}
\newcommand{\loo}{{\lambda}}


\setcounter{secnumdepth}{3}

\renewcommand{\theequation}{S\arabic{equation}}
\renewcommand{\thefigure}{S\arabic{figure}}
\renewcommand{\thesection}{S\arabic{section}}

\section*{\large Supplemental Material}
	
\section{Estimation of Variances}
The experimental results of Fig.~3 are based on an estimate of the variance of the total spin of the ensemble. This section shows how these values can be extracted from the raw data. First, the raw data is presented. Second, the statistical treatment for the unbiased estimation of the underlying variance is described. Finally, we show that the variance of $J_z$ results mainly from number-dependent detection noise. 

\subsection{Measured probability distribution of $J_z$ and $J_{\alpha}$}
The number of atoms in the Zeeman levels is measured by standard absorption imaging with an illumination time of $70\,\mu s$ and an intensity of $40\,\textrm{W}/{\textrm{m}^2}$. The absolute number of atoms was calibrated~{[}40{]} and it was confirmed that shot noise fluctuations are observed for a coherent state [see Fig.~3 (a)]. Without the microwave coupling pulse, a measurement of the number of atoms in the Zeeman levels corresponds to a measurement of $J_z$. While an ideal Dicke state would show no fluctuations at all, we record a finite variance. This finite variance may stem from fluctuations of the number of atoms and from noise in the detection system. Figure~S1 (a) shows the histogram of all measured values for $J_z$ with a total number of atoms between 3000 and 7000. The measured distribution is much narrower than the corresponding result for a coherent state. After a $\frac{\pi}{2}$ microwave pulse, it is possible to record the corresponding histogram in the $J_x$-$J_y$-plane. Since the microwave has an arbitrary phase difference $\alpha$ from the atomic phases, each measurement projects onto a different axis $J_\alpha$ in the $J_x$-$J_y$-plane. The histogram in Fig.~S1 (b) thus includes measurements along all possible directions. The histogram shows super-shot-noise fluctuations, yielding a large effective spin length $J_{\rm eff}.$ The presented data can be used to estimate the second moment of the underlying probability distribution.

\begin{figure}[ht!]
		\includegraphics[width={\columnwidth}]{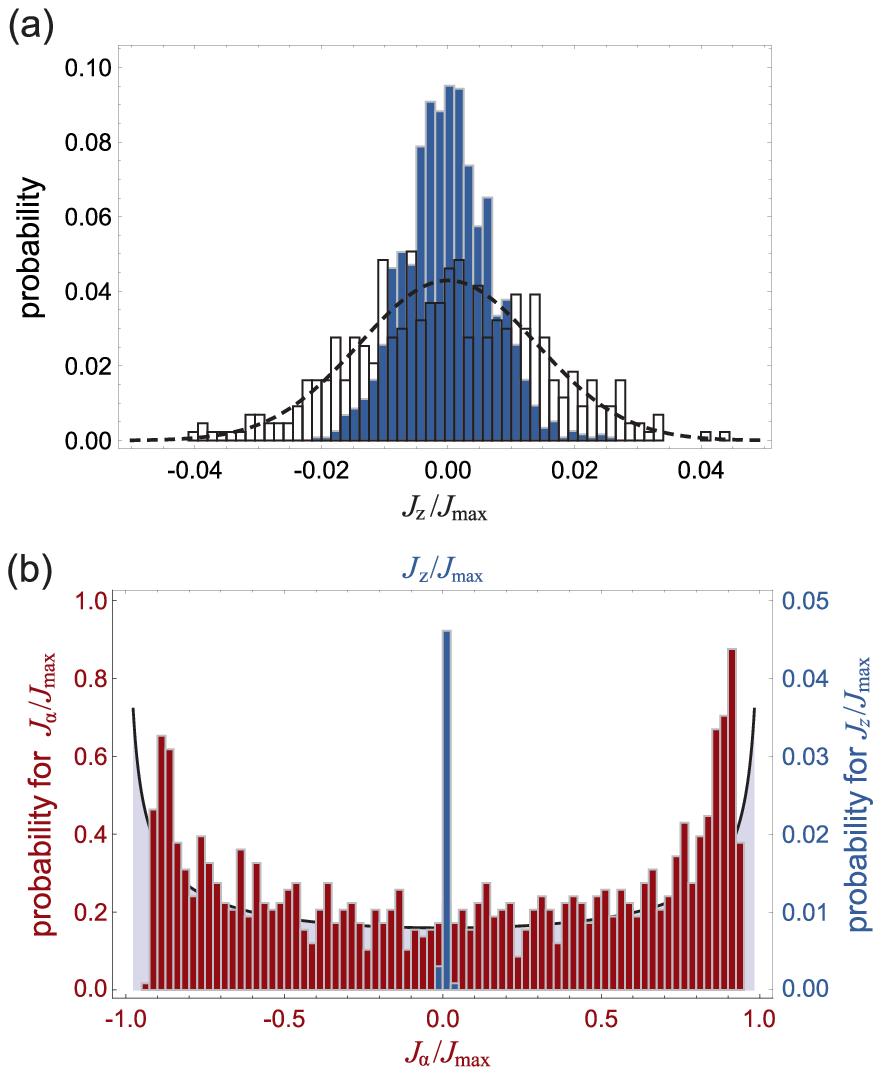}
		\caption{Histograms of the recorded spin measurements. (a) The accumulated measurements of $J_z$  are shown for a Dicke-like state (solid blue columns) and a coherent state (open columns). The distribution of the Dicke-like state is much narrower than the distribution of the coherent state. The latter is very close to a binomial distribution with shot noise fluctuations (dashed line). We corrected for a small offset between the two distributions. (b) The accumulated measurements of $J_\alpha$  are shown for a Dicke-like state (solid red columns). The distribution compares well to the distribution of a perfect Dicke state (solid grey line with shading). It is much wider than the sub-shot-noise distribution of $J_z$ (solid blue columns).}
	\label{figs1}
\end{figure}

\subsection{Unbiased estimation of the second moment of the probability distribution}
The measurement process creates a finite set of random numbers $x_i$ according to a special, non-Gaussian probability function $P(x)$ [see Fig.~S1 (b) as an example]. Such a probability function is well described by its moments $\mu_1=\int{x P(x) dx}$ and $\mu_k=\int{(x-\mu_1)^k P(x) dx}$ for $k\geqslant  2$. The second moment $\mu_2$, which presents the central quantity of interest within our work, can be estimated straightforwardly from the measurements as shown below. However, the variance of this estimate is more difficult to deduce and has previously been gained from split samples~{[}16{]}. In this section, we present a formula for an unbiased estimate of this variance (called second moment variance estimator, SMVE), allowing for the calculation of correct error bars  for the central result of our work [see Fig.~1 (c)].

For a given sample of $n$ independent measurements according to the probability function $P(x)$, it is possible to calculate the sample moments

\begin{eqnarray}
	m_1&=&\frac{1}{n}\sum_{i=1}^n x_i, \nonumber\nonumber\\
	m_2&=&\frac{1}{n}\sum_{i=1}^n (x_i-m_1)^2, \nonumber\\
	m_4&=&\frac{1}{n}\sum_{i=1}^n (x_i-m_1)^4. \nonumber
\end{eqnarray}
The expectation value of $m_2$ is easily calculated to be 

\begin{equation}
\ev{m_2}=\frac{n-1}{n} \mu_2. \nonumber
\end{equation}
It is thus possible to define an unbiased estimator for $\mu_2$:

\begin{equation}
\hat\mu_2=\frac{n}{n-1} m_2. \nonumber
\end{equation}
This estimate shows statistical fluctuations which are described by the variance of $\hat\mu_2$,

\begin{eqnarray}
\rm{var}(\hat\mu_2)
&=& \frac {n^2} {(n-1)^2} \rm{var}(m_2) \nonumber\\
&=& \frac {n^2} {(n-1)^2} \Big( \ev{m_2^2} - \ev{m_2}^2 \Big)\nonumber\\
&=& \frac {n^2} {(n-1)^2} \ev{m_2^2} - \mu_2^2 .
\label{eq:varmu}
\end{eqnarray}
Thus, the problem of finding an estimator for $\rm{var}(\hat\mu_2)$ reduces to finding an estimator for $\mu_2^2$. Hence, we calculate the expectation values $\ev{m_2^2}$ and $\ev{m_4}$ by using augmented and monomial symmetric functions (see Ref.~[50]
p. 416).

\begin{eqnarray}
\ev{m_2^2} &=& \frac {(n-1)^2} {n^3} \mu_4 + \frac {(n-1)(n^2-2n+3)}{n^3} \mu_2^2 \nonumber\\
\ev{m_4} &=&  \frac{n^3-4n^2+6n-3}{n^3} \mu_4 + \frac{3(n-1)(2n-3)}{n^3} \mu_2^2  \nonumber
\end{eqnarray}
This linear system of equations can be solved to yield an estimator for $\mu_2^2$. By substituting this in Eq.~\eqref{eq:varmu}, we obtain the final result for the SMVE,

\begin{equation}
\rm{var}(\hat \mu_2)
= \frac {n}{(n-3)(n-2)} m_4 - \frac{n(n^2-3)}{(n-3)(n-2)(n-1)^2} m_2^2 . \nonumber
\end{equation}
The SMVE allows for a direct calculation of the error bars from the moments of the recorded sample without any assumption on the shape of the probability distribution.

\begin{figure}[t!]
		\includegraphics[width={\columnwidth}]{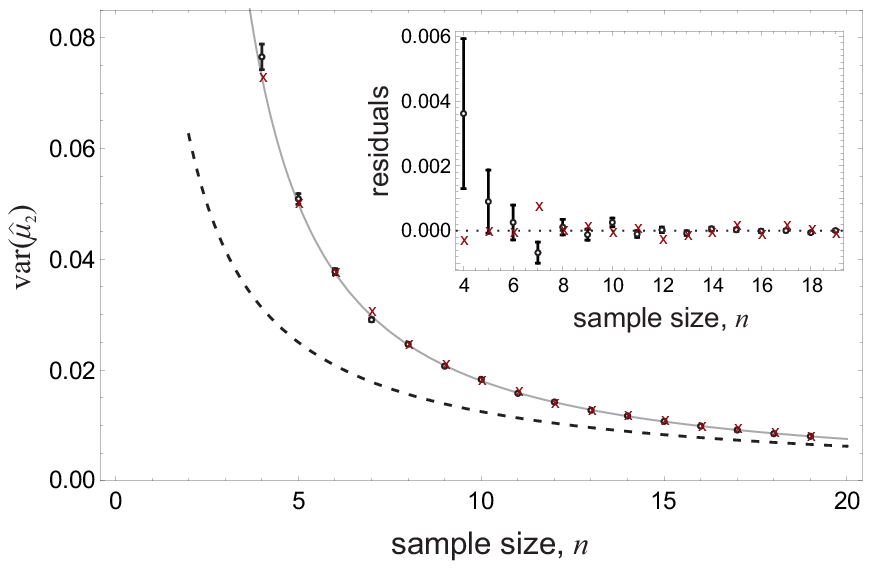}
		\caption{Application of the SMVE to generated random numbers. We have generate random numbers according to a probability function $P(x)$. For each sample size $n$, we have applied the SMVE to $10^4$ samples. The open circles present the mean of the calculated SMVEs with their statistical uncertainties. These results compare well to the directly calculated variance of the $10^4$ sample variances (red solid dots). It is statistically equal to the prediction $\rm{var}(\hat \mu_2)
= \frac {1}{n} \mu_4 - \frac{n-3}{n(n-1)} \mu_2^2$ and completely incompatible with the naive guess $\rm{var}(\hat \mu_2) \approx \frac {1}{n} (\mu_4 - \mu_2^2)$ (dashed line).}
	\label{figs2}
\end{figure}

Figure~S2 shows the result of a Monte-Carlo simulation to demonstrate the application of the SMVE. We generate random numbers according to a probability function $P(x)=\frac{1}{\pi} \sqrt{\frac{1}{1-x^2}}$, similar to Fig.~S1 (b), and accumulate samples of variable size. The SMVE is applied to $10^4$ samples of each size, yielding  estimates for $\rm{var}(\hat \mu_2)$. Figure~S2 shows that these estimates approximate the directly calculated variance of the $10^4$ sample variances very well. It is statistically equal to the prediction gained solely from the shape of the probability distribution.

In summary, the statistical treatment allows for a correct evaluation of the second moment of the underlying probability function and its uncertainty.

\subsection{Estimation of the detection noise}
The second moment gained from the experimental measurements via the statistical treatment above is a combination of the variance $(\Delta J_z)^2$ of the atomic many-particle state and the detection noise. The detection noise comprises an atom-independent part which is dominated by the photoelectron shot noise on the camera pixels and an atom-dependent part. The atom-independent noise was measured continuously during the data acquisition by analysing images without atoms. Since we are interested in an estimate for $(\Delta J_z)^2$, the data in Fig.~3 (a) are corrected for the atom-independent noise. 

\begin{figure}[h!]
		\includegraphics[width={\columnwidth}]{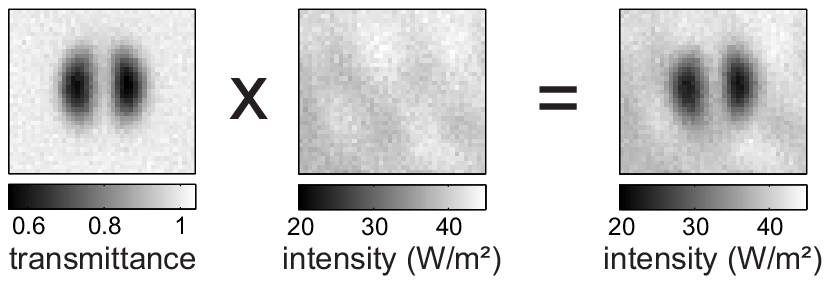}
		\caption{Creation of an artificial absorption image. The optical transmittance of an idealized atomic cloud is calculated from an average of many experimental absorption images. A typical detection image without atoms is multiplied by the optical transmittance to gain a synthetic absorption image with adjustable number of atoms.}
	\label{figs3}
\end{figure}

The atom-dependent detection noise results from fluctuations of the photoelectrons counted on the camera pixels which are stronger at a large number of atoms. Additionally, a change in the number of counted photoelectrons has a larger effect on the estimated number of atoms at high column densities resulting in an increased sensitivity at a large number of atoms. This noise source is not independent of the atomic noise and it is thus not legitimate to subtract it. Nevertheless, we estimate the approximate strength of these fluctuations for comparison with our results. For this purpose, we calculate the mean optical transmittance from many experimental realizations (see Fig.~S3) to approximate an ideal atomic cloud without atom number fluctuations. This optical transmittance image is adjusted to represent clouds with different numbers of atoms. We synthesize absorption images by multiplying empty detection images with the gained transmittance images. These artificial absorption images provide a measure of the atom-dependent detection noise since they do not contain any atom number fluctuations by construction. The resulting estimate for the atom-dependent detection noise is shown in Fig.~3 (a) (dashed line). Although it underestimates the effect of photoelectron shot noise for strongly depleted absorption images, it nevertheless explains the major part of the measured variance $(\Delta J_z)^2$.

\section{boundaries for genuine $k$-particle entanglement}

This section presents a method for the determination of the
entanglement depth based on the measurement of $\langle J_{x}^{2}+J_{y}^{2}\rangle$
and $(\Delta J_{z})^{2}.$  With this method, we determine the allowed regions
for $k$-particle entanglement in Fig.~1 (c). Section~\ref{sub:Simple-numerical-determination}
provides a numerical method to calculate the boundaries. In Sec.~\ref{sub:Proof-for-general},
we present the entanglement criterion with a closed  formula, and we discuss that it
applies to pure states,
mixed states and mixed states with a varying particle number. Finally,
Sec.~\ref{sub:Comparison-with-the} presents a comparison with the
original spin-squeezing criterion of Ref.~{[}24{]}. We show that our
criterion detects a larger entanglement depth for extreme spin-squeezed states in the presence of minimal noise.

\subsection{Numerical determination of the boundaries\label{sub:Simple-numerical-determination}}

The following numerical method can be used to determine the allowed
region in the ($\langle J_{x}^{2}+J_{y}^{2}\rangle$, $(\Delta J_{z})^{2}$)-space
for quantum states with at most $k$-particle entanglement for a given
particle number $N$ [51].
We consider states of
the form
\begin{equation}
\vert\Psi\rangle=\otimes_{n=1}^{M}\vert\psi^{(n)}\rangle,\label{eq:mprodstate}
\end{equation}
where $\vert\psi^{(n)}\rangle$ is the state of the $n$th non-separable subset containing $k_{n}$ qubits and $k_{n}\leqslant  k$. In total, there are $M$ non-separable subsets. Here, {}``qubit'' refers to individual pseudo-spin-$\frac{1}{2}$
atoms in the experiment. We define the collective operators 
\[
J_{l}:=\sum_{n=1}^{M}j_{l}^{(n)}
\]
for $l=x,y,z,$ where $j_{l}^{(n)}$ denotes the components of the $k_{n}$-particle spin operators and act on the $n^{{\rm th}}$ non-separable subset of qubits. Note that we consider $k_{n}=k$ in the main text, whereas here, we extend our discussion to the general case $k_{n}\leqslant  k.$

The total variance $(\Delta J_{z})^{2}$ is given by the sum of the variances of the $k_{n}$-particle spin operators
\begin{equation}
(\Delta J_{z})^{2}=\sum_{n}(\Delta j_{z}^{(n)})^{2}.\label{eq:varJz}
\end{equation}
On the other hand, for a state of the form \eqref{eq:mprodstate} 
\begin{align*}
\langle J_{x}^{2}+J_{y}^{2}\rangle & =\sum_{n}\langle\left(j_{x}^{(n)}\right)^{2}+\left(j_{y}^{(n)}\right)^{2}\rangle\\
 & +\sum_{m\ne n}\left(\langle j_{x}^{(m)}\rangle\langle j_{x}^{(n)}\rangle+\langle j_{y}^{(m)}\rangle\langle j_{y}^{(n)}\rangle\right).
\end{align*}
Since for non-negative values $\{x_{l}\}_{l=1}^{L}$ and positive
integer $L$ we have
\[
\sum_{l\ne m}x_{l}x_{m}\leqslant (L-1)\sum_{l}x_{l}^{2},
\]
we obtain 
\begin{align}
\langle J_{x}^{2}+J_{y}^{2}\rangle\text{\ensuremath{\leqslant}} & \sum_{n}\langle\left(j_{x}^{(n)}\right)^{2}+\left(j_{y}^{(n)}\right)^{2}\rangle\nonumber \\
+ & \left(M-1\right)\sum_{n}\left(\langle j_{x}^{(n)}\rangle^{2}+\langle j_{y}^{(n)}\rangle^{2}\right).\label{eq:Jx2Jy2_ineq}
\end{align}

For simplicity, we assume that $N$ is divisible by $k.$ In this
case, states of the form 
\begin{equation}
\vert\Psi\rangle=\vert\psi\rangle^{\otimes\frac{N}{k}}\label{eq:phialpha}
\end{equation}
saturate the inequality \eqref{eq:Jx2Jy2_ineq}, where $\vert\psi\rangle$
is a $k$-qubit state. Due to convexity arguments, it is sufficient to look for states of the form~\eqref{eq:phialpha} to calculate the boundary points. A boundary point can be obtained for a given $X=(\Delta J_{z})^{2}$
from
\begin{align}
\langle J_{x}^{2}+J_{y}^{2}\rangle(X)= & \max_{\vert\Psi\rangle,\frac{N}{k}(\Delta j_{z})^{2}=X}\bigg[\tfrac{N}{k}\langle j_{x}^{2}+j_{y}^{2}\rangle{}_{\vert\text{\ensuremath{\psi\rangle}}}\nonumber \\
+ & \left(\tfrac{N}{k}-1\right)\tfrac{N}{k}\left(\langle j_{x}\rangle_{\vert\text{\ensuremath{\psi\rangle}}}^{2}+\langle j_{y}\rangle_{\vert\text{\ensuremath{\psi\rangle}}}^{2}\right)\bigg].\label{eq:boundary}
\end{align}

Thus, a constrained optimization for a given ($\Delta j_{z})_{\vert\psi\rangle}^{2}$
over $\vert\psi\rangle$ has to be performed. This can be simplified
further as follows. For even $k,$ the states at the boundary can
be sought in the form~\eqref{eq:phialpha}, where $\vert\psi\rangle$
is the ground state of the spin-squeezing Hamiltonian
\begin{equation}
h(\lambda)=j_{z}^{2}-\lambda j_{x}.\label{eq:ham_boundary}
\end{equation}
Thus, an optimal state $\vert\psi\rangle$ is obtained from
spin squeezing {[}13{]}. Note that the ground state of $h(0)$ is degenerate.
In this case, the symmetric ground state has to be chosen, i.e., the symmetric Dicke state with $\langle j_{z}\rangle=0.$

Hence, the boundary points can be obtained for even $k$ as a function
of a single real parameter $\lambda$ as
\begin{align*}
\langle J_{x}^{2}+J_{y}^{2}\rangle(\lambda)= & \bigg[\tfrac{N}{k}\langle j_{x}^{2}+j_{y}^{2}\rangle{}_{\vert\text{\ensuremath{\psi\rangle}(\ensuremath{\lambda}})}\\
+ & \left(\tfrac{N}{k}-1\right)\tfrac{N}{k}\left(\langle j_{x}\rangle_{\vert\text{\ensuremath{\psi\rangle}(\ensuremath{\lambda}})}^{2}+\langle j_{y}\rangle_{\text{\ensuremath{\vert\psi\rangle}(\ensuremath{\lambda}})}^{2}\right)\bigg],\\
(\Delta J_{z})^{2}(\lambda)= & \tfrac{N}{k}(\Delta j_{z})_{\text{\ensuremath{\vert\psi\rangle}(\ensuremath{\lambda}})}^{2},
\end{align*}
where $\vert\psi\rangle(\lambda)$ is the ground state of $h(\lambda).$
This also means that states of the form $\vert\psi\rangle^{\otimes\frac{N}{k}}(\lambda)$
correspond to points on the boundary. Since $\langle j_{z}\rangle_{\vert\psi\rangle(\lambda)}=0,$
we have $\langle J_{z}\rangle=0$ for the states on the boundary mentioned
above. Any state beyond the boundary is at least $(k+1)$-particle
entangled. 

\subsection{Proof for general states with a large number of particles \label{sub:Proof-for-general}}

In the previous section, we have presented a numerical method to calculate
the boundary for $k$-particle entangled states assuming that the
state is a tensor product of $k$-qubit pure states and the particle
number is fixed. It is possible to prove that these boundaries are
valid for general states \eqref{eq:mprodstate} with $k_{n}\leqslant  k.$

To obtain a closed formula for the boundary, we employ the definition
{[}13{]} 
\begin{align*}
F_{j}(X):=\tfrac{1}{j} & \min_{\frac{\langle j_{x}\rangle}{j}=X}(\Delta j_{z})^{2}.
\end{align*}
The spin-squeezing criterion for $k$-particle entangled states is
given as
\begin{equation}
(\Delta J_{z})^{2}\geqslant J_{\max}F_{\frac{k}{2}}\left(\frac{\sqrt{\langle J_{x}\rangle^{2}+\langle J_{y}\rangle^{2}}}{J_{\max}}\right).\label{eq:extreme}
\end{equation}
Equation~\eqref{eq:extreme} is valid for any tensor product of states of the form \eqref{eq:phialpha} with $k_n\le k$ {[}13,S3{]}.

Moreover, for pure $k$-particle entangled states it is straightforward to show that
\begin{equation} 
\langle J_{x}^{2}+J_{y}^{2}\rangle\leqslant  J_{\max}(\tfrac{k}{2}+1)+\langle J_{x}\rangle^{2}+\langle J_{y}\rangle^{2}.\label{eq:bound3}
\end{equation}
Hence, using the properties of $F_{j}(X),$ for states with $k$-particle entanglement,
\begin{align} 
(\Delta J_{z})^{2}\geqslant
J_{\max}F_{\frac{k}{2}}\left(
	\frac{\sqrt{\langle J_{x}^{2}+J_{y}^{2}\rangle
		-J_{\max}(\tfrac{k}{2}+1)}}
	{J_{\max}}
\right)\label{eq:ineq}
\end{align}
holds. Naturally, we can use the formula only if the expression under
the square root is positive. Otherwise, the lower bound on $(\Delta J_{z})^{2}$ is trivially zero. For large $N$ and $k\ll N,$ the first
term under the square root in Eq.~\eqref{eq:ineq} is $\sim N^{2},$
while the second one is $\sim N.$ Thus, we obtain approximately 
\begin{equation} 
(\Delta J_{z})^{2}\gtrsim J_{\max}F_{\frac{k}{2}}\left(\frac{\sqrt{\langle J_{x}^{2}+J_{y}^{2}\rangle}}{J_{\max}}\right).\label{eq:largeN}
\end{equation}
Note that, since $F_j(x)\leqslant  \frac{1}{2},$ a sub-Poissonian variance, i.e., $(\Delta J_{z})^{2}<\frac{N}{4}$ is required to detect multi-particle entanglement.

The inequality~\eqref{eq:ineq} can be used to quantify the entanglement depth of pure states. It
gives the same boundary for $k$-particle entangled states as the method of Sec.~\ref{sub:Simple-numerical-determination}.
It can also be shown that our criterion holds not only for pure states,
but also for general mixed states [52].
Moreover, it can be generalized
to the experimentally important case of mixed states with a fluctuating
total number of particles. Since the total proof exceeds the scope
of this publication, it will be published elsewhere~[53].

\subsection{Comparison with the spin-squeezing criterion \label{sub:Comparison-with-the}}

Our criterion reliably detects the entanglement depth of Dicke states. In particular, it detects the symmetric Dicke state with $\langle J_{l}\rangle=0$ for $l=x,y,z$ as fully $N$-particle entangled, since the inequality  \eqref{eq:ineq} with $k=N-1$ is violated. In this section, we show that our criterion is also valuable for the evaluation of spin-squeezed states, since it outperforms the criterion of Ref.~{[}24{]} in the presence of noise.

\begin{figure}
\centering{\includegraphics[width=6.5cm]{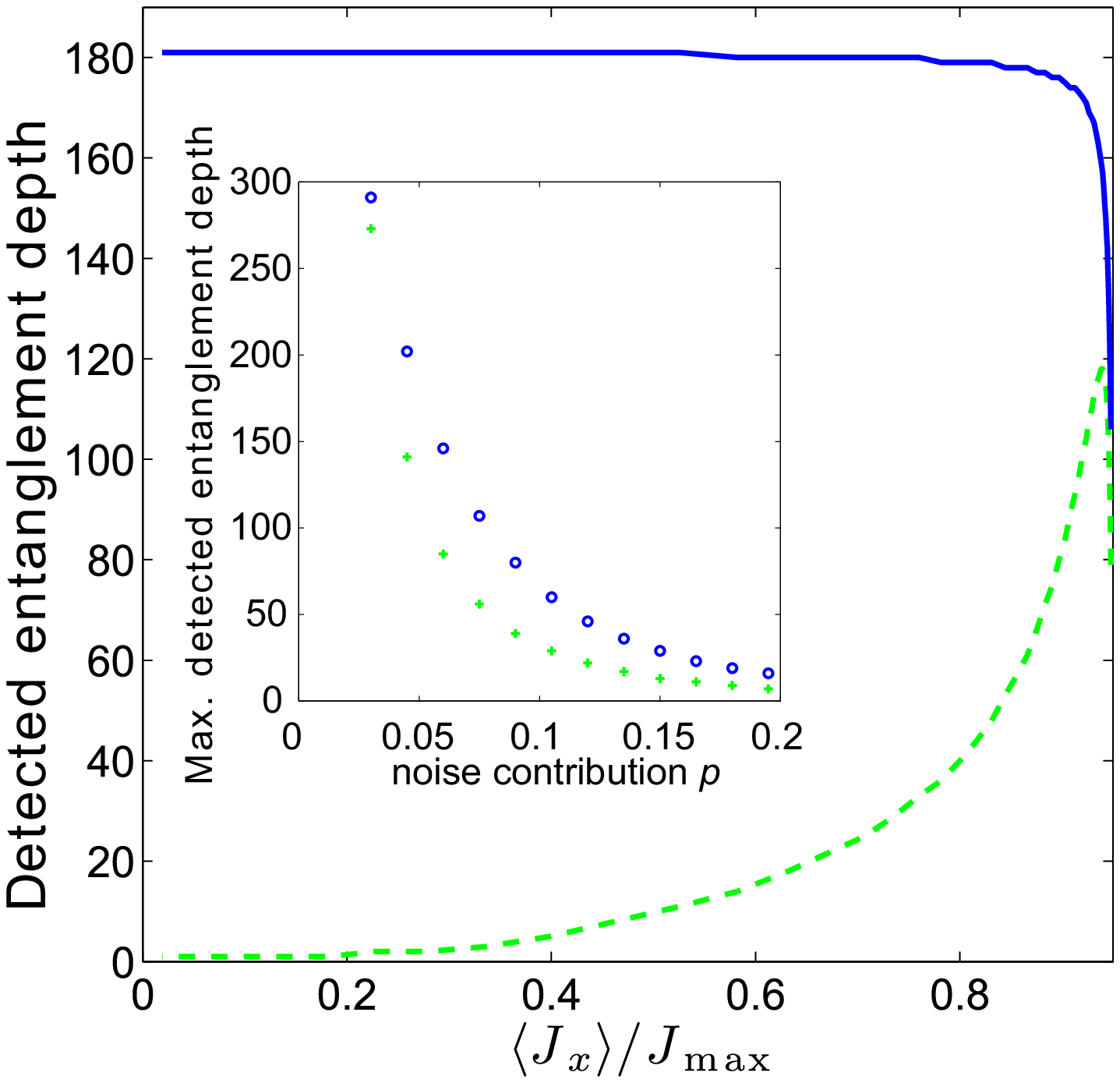}\caption{Comparison with the spin-squeezing criterion. The graph shows the entanglement depth detected by the condition~\eqref{eq:ineq} (solid line) and the spin-squeezing condition~\eqref{eq:extreme} (dashed line) for $N=4000$ spin-$\frac{1}{2}$ particles with additive white noise to account for imperfections. For states that are not
completely polarized, Eq.~\eqref{eq:ineq} detects a considerably larger entanglement
depth. The inset shows that the maximal detected entanglement depth depending
on the noise contribution is larger for our criterion (circles) than
for the spin-squeezing criterion (crosses) if some very small noise is
present.} \label{fig:kent}}
\end{figure}

In order to compare the performance of the two criteria, we consider the ground states of the spin-squeezing Hamiltonian
\begin{equation}
H(\Lambda)=J_{z}^{2}-\Lambda J_{x},\label{eq:hground}
\end{equation}
for $N=4000$ spin-$\frac{1}{2}$ particles. For $\Lambda=\infty,$
the ground state is fully polarized. For $\Lambda=0,$ it is the symmetric
Dicke state. In principle, such states are detected by the spin-squeezing criterion of Ref.~{[}24{]} as fully $N$-particle entangled for all $\Lambda>0$. However, this statement only holds for ideal pure states. In experimentally realistic situations, small noise contributions are always expected, especially for the case of large numbers of particles as considered here. While the criterion of Ref.~{[}24{]} becomes extremely sensitive to noise for strongly squeezed states, our criterion is much more robust.

We account for these small noise contributions by mixing the density matrix of the ideal spin-squeezed state $\rho_\mathrm{id}$ with a noisy state $\rho_\mathrm{n}$. The noisy state is chosen such that each atom is in an incoherent 50/50 mixture of its two spin states. For a quantitative comparison, we estimate the entanglement depth of the state $\rho = (1-p) \, \rho_\mathrm{id} + p \, \rho_\mathrm{n}$ with a noise contribution of $p=0.05$. Fig.~\ref{fig:kent} shows the detected entanglement depth for the spin-squeezing criterion \eqref{eq:extreme} and our criterion \eqref{eq:ineq}. For strongly squeezed states, where $\ev{J_x} \ll J_{\max}$, our criterion detects a large entanglement depth, while the result of the method described in Ref.~{[}24{]} tends to zero. The robustness against noise exhibited in this example is a general property and is independent of the exact type of noise. 
 
 In summary, our criterion detects the entanglement depth of both spin-squeezed states and more general states in experimentally realistic situations. Most prominently, it is ideally suited for the characterization of Dicke states, as produced in our experiments.

\end{document}